\documentstyle[prd,aps,epsf]{revtex}  
\begin{document}            
\draft
\twocolumn[\hsize\textwidth\columnwidth\hsize\csname 
@twocolumnfalse\endcsname

\title{The surface brightness of dark matter: \\ 
unique signatures of neutralino annihilation in the Galactic halo.} 

\author{Carlos Calc\'{a}neo--Rold\'{a}n\footnotemark[1] \&
Ben Moore\footnotemark[2].}
\address{Department of Physics, Durham University, 
Science laboratories, Durham, DH1 3LE, UK.}

\date{\today}
\maketitle  
\begin{abstract} 
We use high resolution numerical simulations of the formation of cold
dark matter halos to simulate the background of decay products from
neutralino annihilation, such as gamma-rays or neutrinos.  Halos are
non-spherical, have steep singular density profiles and contain many
thousands of surviving dark matter substructure clumps. This leads to
several unique signatures in the gamma-ray background that may be
confirmed or rejected by the next generation of gamma-ray experiments.
Most importantly, the diffuse background is enhanced by over two
orders of magnitude due to annihilation within substructure halos. The
largest dark substructures are easily visibly above the 
background and may account for the unidentified EGRET sources. A deep
strip survey of the gamma-ray background would allow the shape of the
Galactic halo to be quantified.
\end{abstract}

]


\footnotetext[1]{Email:C.A.Calcaneo-Roldan@durham.ac.uk}
\footnotetext[2]{Email:Ben.Moore@durham.ac.uk}

\section{Introduction}
\label{s:Intro}

Determining the nature of dark matter is of fundamental importance to
both Astronomy and Particle Physics. Both theory and observational
data currently favour a universe with a matter density that is
dominated by non-baryonic particles. Many candidates have been
proposed: some are known to exist, others are more speculative (e.g.,
Ref. ~\cite{ellis98} and references therein).  Structure formation in
a universe dominated by cold dark matter (CDM) has been extensively
tested against observations and the model has proven highly successful at
reproducing the large scale properties and distributions of galaxies
~\cite{davis85,baugh}. On the non-linear scales
of galactic halos it remains to be confirmed wether the model can
successfully reproduce the observational data
~\cite{moore94,moore99a,Klypin99}.  

Direct detection in the laboratory is the ultimate technique for
verifying the existence of dark matter particles (see, Ref. 
~\cite{collar99}). However, even the
most popular candidate for dark matter, the neutralino, has a cross
section that spans many orders of magnitude and the current laboratory
searches are only just becoming sensitive to the cosmologically
interesting parameter range. Presently, Astronomical
observations provide the best insights into the nature of the dark matter,
furthermore 
direct detection relies on the existence of a smooth component of dark
matter.

Within the next few years indirect detection of neutralinos will
provide interesting constraints on their possible cross-section and
masses.  Neutralino-neutralino annihilation produces observable
photons (as well as a host of other particles) that may be observed as
a diffuse gamma-ray background from the halo surrounding the Milky Way
as discussed in Refs.
~\cite{gunn78,stecker78,silkS84,turner86,silkB87,bouquet89,lake90}, 
and more recently, in Refs.
~\cite{berg2,berg98a,baltz99,gondolo99a,baltz00}.

Renewed interest in these predictions has recently arisen because of 
an unexplained component of diffuse high energy photons in the Egret data
(e.g. Ref.~\cite{dixon98}), and also the possibility of an excess from the
center of the Galaxy itself ~\cite{Mayer98}, unexpected clumpy emission 
and the unresolved ``discrete sources''
~\cite{hartman}. Progress in this area will result from several new and
sensitive gamma-ray surveys such as GLAST ~\cite{glast} and 
VERITAS ~\cite{veritas}.

The efficiency of the annihilation process is strongly dependent on the local
density and the cross-section of the neutralino.  Many authors have
calculated the expected flux from the Galactic halo using simple
models for the expected mass distribution of neutralinos within the
Galaxy (see, e.g., Refs.~\cite{berg98a,gondolo99a}) or from
its satellites ~\cite{lake90,baltz00}.  

Advances in computational Cosmology have lead to several
recent breakthroughs that have direct relevance to the detection
of dark matter. In particular, the
numerical resolution that can be achieved using parallel
computational techniques is now sufficient to study the 
internal structure
of dark matter halos  that form within a 
cosmological context. The results of these simulations have important
implications for indirect (and direct) detection of dark matter
candidates. Most significantly for particle-particle annihilation, we
are now confident that the central density profile of CDM halos follows a
singular power law down to small scales 
~\cite{carlberg94,navarro96,moore98a,ghigna98}. 
Thus we may expect a point like source of mono-chromatic
gamma-rays emanating from the center of the Milky Way, where the annihilation 
rate will be very high. 

A second fundamental prediction of the CDM model is that previous
generations of the merging hierarchy survive within halos
~\cite{ghigna98}.  Halos that accrete into larger systems may be
tidally stripped of most of their mass, however their dense central
cores survive and continue to orbit within the parent halos.  This may
present some problems for the CDM model since the predicted number of
satellites within the Milky Way's halo is 50--100 times as many as
observed ~\cite{moore99a}. If the CDM model is correct, then
only a fraction of these satellites must have formed stars and most of
the substructure remains as dark objects within the Galactic halo.

The possibility of an enhanced gamma-ray background from dark matter
substructure was explored by Bergstr\"om {\it et al.}~\cite{berg99}, 
who made simple
assumptions as to the mean density and abundance of such clumps.  We
can now use the high resolution N-body simulations to directly measure
these quantities.  The simulations also allow us to study the
influence of the halo shapes on the diffuse gamma-ray background and
the intensity of the central halo emission that arises from the
singular dark matter density profiles.  This paper is organized as
follows. In Section~\ref{s:estimating} we explore the gamma-ray background 
that results from the smoothly distributed component of dark matter using
both analytical and simulated halos.
In Section~\ref{s:subs} we focus on the substructure 
within halos. Our conclusions are summarised in Section~\ref{s:concs}.

\section{The sky distribution of the gamma-ray background}
\label{s:estimating}

In what follows we will consider a flux of photons (or other particles) 
that are a by-product of
the annihilation of dark matter particles within the smooth component of
dark matter that surrounds the Galaxy.  
It is not our intention to discuss the details of
neutralino interactions, a complete overview on these processes 
(and super-symmetric matter in general) can be found in 
Refs.~\cite{jungman96,berg1}.

\subsection{Model neutralino halos}
\label{sub:theocurve}

We calculate the gamma-ray flux along a given line of sight through 
a spherically symmetric galactic halo using: 
\begin{equation}
{\phi(\psi)=\frac{K}{4\pi} \int_{Line \; of \; sight} \rho^2(l)dl(\psi)}
\label{eq:1}
\end{equation}
where $\psi$ is the angle between the direction of galactic center and 
observation; $\rho$, the density of dark matter at distance $l$ from the
observer. We have summed up the dependence of 
the flux on neutralino mass and interaction cross section in the constant $K$.
This is enough scope for the present discussion - it is straightforward
to take our results and input a neutralino cross-section, $<\sigma v>$, and 
mass, $M_{\chi}$ to determine the absolute gamma-ray flux (where $K$ is  
defined to be $<\sigma v> / M^{2}_{\chi}$). Our results can also be used to 
infer the sky distribution of other products of the annihilation, such as 
neutrinos or positrons.

The line of sight distance, $l$, is related to the radial distance from the 
halo center, $r$, via 

\begin{center}
$r^2 = l^2 + R_o^2 -2lR_o\cos(\psi)$
\end{center}   

\noindent where $R_o$ is our galacto-centric distance, taken here to have
the IAU standard value of $R_o = 8.5$kpc~\cite{kerr86}, and 
$\psi$ is related to galactic coordinates ($\ell$, $b$) through

\begin{center}
$\cos(\psi)= \cos(\ell)\cos(b)$.
\end{center}   
    
For the halo density profile, $\rho(r)$, we take the latest results from 
the highest resolution numerical simulations of galactic halos carried 
out to-date~\cite{moore99a}.  
These authors simulated 6 different galactic mass halos with force
resolution of $0.5$ kpc and mass resolution of $10^6M_\odot$. (Throughout 
the paper we will use the Hubble constant value of 
$H_o = 100 h {\rm kms}^{-1}{\rm Mpc}^{-1}$ and $h=0.5$; as adopted for the simulations.)
The best fitting
density profile to these data is (subscript {\it moore}):
\begin{equation}
{\rho_{moore}(r) = \frac{\rho'_{moore}}{(r/a)^{1.5}(1+(r/a)^{1.5})}
\label{eq:0}}.
\end{equation} 
Where $r$ is the distance from the halo center, $a=r_{200}/c_{moore}$ the scale 
radius for halos of mass $\approx 1\times10^{12}M_\odot$.
The virial radius of our fiducial Galactic halo, $r_{200}\approx 300$ kpc,
is defined as the radius of a sphere at which the mean overdensity is 200
times the cosmological mean density. (A central density profile of slope -1.5 
on galactic scales was also found by Jing \& Suto \cite{Jing} and confirmed 
as an asymptotic slope by Ghigna {\it et al.} \cite{ghigna2000}.)

We also compare this profile with that determined by 
Navarro {\it et al.}~\cite{navarro96}
using a sequence of lower resolution studies (subscript {\it nfw}) 
(the main difference
being that the central dark matter density profile has a slope of -1):
\begin{equation}
{\rho_{nfw}(r) = \frac{\rho'_{nfw}}{(r/a)(1+r/a)^2}}
\end{equation} 
and the modified isothermal profile with a constant density core 
(subscript {\it is}):
\begin{equation}
{\rho_{is}(r) = \frac{\rho'_{is}}{[1+(r/a)^2]^{3/2}}}.
\end{equation}  

The scale radius, $a$, is determined directly from the numerical
simulations, except for the modified isothermal model which
we normalise to match the observational rotation curve data 
(as in Ref.~\cite{kravstov98});
$a_{is} = 24.3$ kpc, $a_{nfw} = 27.7$ kpc and $a_{moore} = 33.2$ kpc 
(this radius is directly related to the concentration parameter, 
$c = {\rm r}_{200}/a$).
We normalise each density profile such that the peak circular velocity,
$v_{peak} = 200$ km/s (the maximum of the $v_c=\sqrt{GM/r}$ curve), which gives:
$\rho'_{is} = 4.96\times10^6{\rm M_{\odot}}$ kpc$^{-3}$, 
$\rho'_{nfw} = 5.11\times10^6{\rm M_{\odot}}$ kpc$^{-3}$ and 
$\rho'_{moore} = 1.64\times10^6{\rm M_{\odot}}$ kpc$^{-3}$. 
We plot the effective circular 
velocity profiles and density profiles of these model halos in
Fig.~\ref{f:001}(a) and Fig.~\ref{f:001}(b) respectively.

In Fig.~\ref{f:002} we plot the flux, $\phi$, along the line of
sight through a spherical Milky Way halo using the above density profiles
as the observer looks towards the
Galactic center at $\psi=0^{\circ}$, to the Galactic anticenter 
at $\psi=180^{\circ}$. As expected, the central annihilation
flux depends strongly on the form of the inner density profile.
At an angle of five degrees from the Galactic center, the ratio of 
fluxes from the three different profiles, 
{\it moore:nfw:is} are 1000:100:1. 

The peak central value depends upon the distance from the Galactic center
that we are willing to consider integrating from \--- the flux slowly 
diverges for the density profile in Eq.~(2).
However, within a given radius, most of the neutralinos would have self
annihilated leaving a tiny constant density core.
We can estimate the size of this core using
${(n \sigma v)}^{-1} = t_h$, where $t_h\sim 10$ Gyrs is the Hubble time.
Taking a typical cross section, $\sigma v =10^{-30}{\rm cm}^3$ s$^{-1}$, 
and adopting the Moore {\it et al.} density profile we find that the 
annihilation radius within the Milky Way is approximately $4 \times 10^{-7}$ 
parsecs $\approx 10^{-12}r_{200}$. 

The total flux that arises within 5 degrees of the Galactic 
center using the Moore {\it et al.} density profile is a factor of 20 
larger than that found using the NFW profile (both integrated down to the 
annihilation radius calculated above).

\subsection{Comparison with high resolution CDM simulations}
\label{sub:simulations}

We can use the numerical simulations to compare directly with the above 
predictions
that were obtained assuming spherical symmetry. 
We refer the reader to Moore {\it et al.} (Ref. ~\cite{moore99a}) for 
details of the numerical
simulations.\footnote[1]{Images, data and movies of these dark matter 
simulations can be downloaded from http://www.nbody.net.}
To construct the expected gamma-ray sky maps we
choose a simulated dark matter halo at a redshift z=0 that has a peak
circular velocity of $\sim 200 {\rm km s^{-1}}$ and a total mass, within 
the virial radius, $r_{200}=300 {\rm kpc}$, of 
$1 \times 10^{12} {\rm M_{\odot}}$. This simulated halo is from the 
``Local Group''
simulation and is close to our fiducial Milky Way cold dark matter halo that
we adopted in the previous section. 

N-body simulations attempt to simulate a collisionless fluid of dark matter
using discrete massive particles. We calculate the local density at the
position of each particle by averaging over its nearest 64 neighbours.
The observer is placed 8.5 kpc from the halo center (defined using the
most bound particle in the simulation) and we sum up the flux
of annihilation products along each line of sight using the 
discrete equivalent to Eq.~(~\ref{eq:1}):

\begin{equation}
{\Phi(\ell,b)=\frac{K}{\Omega}\sum_{L\;O\;S}\rho_{i}^2(\ell,b)
\Delta r_{i}(\ell,b)
\label{eq:4}}
\end{equation}
where $\ell$, $b$ are galactic longitude and latitude respectively.
The flux is binned in angular windows of size 
$\Omega = 1^{\circ} \times 1^{\circ}$ and in the radial 
direction in fixed increments $\Delta r_{i} =$ 1kpc.

The simulated dark matter halos are typically flattened oblate or prolate 
systems~\cite{barnes87}. 
We do not know a-priori in which axis the stellar disk would be located, 
therefore
we show two all-sky maps using the same dark matter halo
but viewed using two different locations for the ``observer'':
Figure~\ref{f:003}(a) and Fig.~\ref{f:003}(b)
has the observer located on the short and long axes respectively.
Both of these plots show the enhanced brightening towards the halo
center, as well as some clumpy substructure in the halo itself. Note
that both the central halo and the centers of the substructure halos
are artificially ``dimmed'' in these plots due to the numerical
resolution $\sim 0.5$ kpc, which sets a maximum density that can be
resolved.  The non-spherical shape of the halo is also clearly evident 
by inspecting the plots with different observer positions.

Recent estimates for the shape of the Milky Way's halo (see, e.g., 
~\cite{olling99} and references therein), suggest
that it may be flattened with a short/long axis ratio of 0.5.  An independent
estimate from the orbit of the Sagittarius debris stars yields a nearly 
spherical dark matter halo ~\cite{ibata}. The
simulated halo that we have chosen to analyse represents a typical
prolate CDM halo with a short to long axis ratio of 0.4, and
intermediate to long axis ratio of 0.5.

It is straightforward to estimate the effects of flattened dark matter
halos by modifying Eq. (~\ref{eq:1}) to accommodate triaxial shaped
bodies. The simplest way to achieve this is to change from spherical 
coordinate $r$ to
\begin{center}
$\xi^2 = \frac{x^2 + y^2}{b^2} + \frac{z^2}{c^2}$
\end{center}
\noindent where $b > c$ for the oblate case, and  $b < c$ for prolate
and we leave $z$ as the axis of symmetry. A 2d visualization of these 3d 
shapes is illustrated in Fig.~\ref{f:004}. 

In Fig.~\ref{f:005} we plot spherical, oblate (2:1) and prolate 
(2:1) versions of the integral 
in Eq. (~\ref{eq:1}) using the Moore {\it et al.}
(1999) density profile.  The observer is located on a plane parallel
to the axis of symmetry, again at a distance $R_o = 8.5$kpc from
the center of the halo. The halo shape leads to little
difference towards the Galactic center, but at the anti-center prolate
halos can be 100 times brighter than oblate halos.
 
We can also compare the predicted angular flux with that measured
directly from the N-body simulation.  The annihilation flux is
averaged in ten degree bins from the simulated dark matter halo, along
a great circle from from the galactic center to its anti-center.  This
direct measurement of the flux is also plotted (as points) in
Fig.~\ref{f:005}. These data are particularly noisy due to the
large numbers of substructure clumps in the simulation - the spike at
$\psi=125^{\circ}$ is due to a massive dark clump that happens to lie exactly 
along this chosen line of sight.

\section{Substructure}
\label{s:subs}

\subsection{Enhancement of global flux due to substructure}

Cold dark matter substructure clumps have singular density profiles that will
be a significant source of annihilation products.  The velocities and spatial
distribution of dark matter substructure is unbiased with respect to the
smooth dark matter background~\cite{ghigna2000}.  Therefore, to first order, 
substructure increases the global sky brightness in any given direction.  However, 
the details depend on how much substructure survives within the solar radius
and also on how far down the mass function substructure halos form and 
survive.

First we will estimate the annihilation flux from clumps of dark matter that
are known to exist in the Galactic halo {\it i.e.}  the dark matter halos that
surround the Magellanic Clouds and dwarf spheroidal galaxies. In fact, 
high-Energy gamma-ray emission from the Large Magellanic Cloud (LMC) was 
detected with EGRET by Sreekumar {\it et al.}~\cite{sreekumar} in 1992 
(although the origin of this emission was reported to be the interaction of 
cosmic rays with interstellar matter). 

We estimate the average flux, $\Phi_{AV}$, from the dark matter halos that 
surround some of the principal structures in the Local Group: The ``Andromeda 
Galaxy''; M31 ($v_{peak}=200{\rm km s^{-1}}$ at a distance of 700 kpc), The 
Large and Small Magellanic Clouds ($v_{peak}=70{\rm km s^{-1}}$ 
and $v_{peak} = 40{\rm km s^{-1}}$, respectively; both at a distance of 50 
kpc), Draco ($v_{peak} = 10{\rm km s^{-1}}$ at a distance of 50 kpc) and a 
small dark matter clump ($v_{peak} = 2{\rm km s^{-1}}$ at a distance of 
10 kpc). A sketch of the geometry is given in Fig.~\ref{f:006}.

The total flux from a substructure halo at distance $R_c$ from the observer
is
\begin{equation}
{\Phi_{TOT}(R_c)=\frac{K}{R_c^2} \int \rho^2(r)r^2dr.}
\label{eq:rflux}
\end{equation} 
by considering the central $\Delta\Omega = 1^{\circ} \times 1^{\circ}$ patch 
over each clump, we define the maximum integration limit in~\ref{eq:rflux} above and 
the average flux is then 
\begin{equation}
{\Phi_{AV}=\frac{\Phi_{TOT}}{\Delta\Omega}}
\label{eq:aveflux}
\end{equation}
(we set $\Delta\Omega$ in steradians, so we may compare directly  with the 
smooth flux of Section~\ref{s:estimating}).

For the dark matter distribution within the substructure clumps we use
the Moore {\it et al.} profile, which provides a good fit to the
smallest, well-resolved substructure halos. The concentration of CDM
halos is a function of mass~\cite{klypin99b} and for the density profile
in Eq.~\ref{eq:0} this can be written:
\begin{equation}
{c_{moore} \approx 102(\frac{M_{\it vir}}{1h^{-1}M_{\odot}})^{-0.084}}.
\label{eq:consrule}
\end{equation}   
This defines the scale radius of each substructure clump: 
$a_{\rm M31} = 33.3{\rm kpc}, \ 
a_{\rm LMC} = 6.7{\rm kpc}, \ 
a_{\rm SMC} = 3.1{\rm kpc}, \ 
a_{\rm Draco} = 0.5{\rm kpc}, \ 
a_{\rm Tiny} = 0.05{\rm kpc} $.

The integral in Eq.~\ref{eq:rflux} diverges as $r \rightarrow 0$ for the
density profile that we are using, however, even the smallest substructure
halos will have a maximum density set by the radius within which most of the
neutralinos would have self annihilated. We therefore present results for the 
average flux from these clumps as a function of the minimum integration 
radius ${\rm R_{min}}/a$ in Fig.~\ref{f:007}, where $a$ is the scale radius 
as defined above.

For comparison, we plot the range of background emission at the
Galactic anti-center as the shaded line in Fig.~\ref{f:007}. The Tiny clump 
is only marginally visible above the background flux (depending on whether or
not the Galactic halo is prolate or oblate) whereas most of the subhalos
are easily visible. Also for comparison we have plotted the flux from the
inner region of the galaxy which is the brightest of these sources.

Although the Galactic halo is expected to 
contain just a few clumps more massive that the Magellanic 
Clouds, there are many thousands of smaller mass objects. 
The mass function of substructure is a power law close to 
$dn(m)/dm \propto m^{-1.9}$ or in terms of circular velocity 
$dn(v_c)/dv_c \propto v_c^{-3.8}$ (~\cite{ghigna2000}).
Above a circular velocity  $v_{peak}=10$ km s$^{-1}$ and $1$ km s$^{-1}$ we
expect the galactic halo to host roughly 1000 and $5 \times 10^5$ substructure
halos respectively. Future simulations should be able to measure how far
down the mass function substructure halos can survive as well as to determine
their central density profiles.
(We note that the highest resolution simulation to date resolved the
substructure within a dark matter mini-halo of mass $10^7M_\odot$.
The force resolution was 10 parsecs and the mass resolution was 10M$_\odot$
allowing substructure with peak circular velocities as low as a few 
hundred meters per second to be resolved. The survival of substructure 
continues even down to this scale, where the slope of the power spectrum 
is close to -3.)

We calculate the total flux from substructure using Monte-Carlo
techniques. First we generate a list of peak circular velocities and
positions of $5\times10^5$ substructure halos in the range of
1\---70km s$^{-1}$.  (Distances are randomly selected using the Moore
{\it et al.} density profile and peak circular velocities are randomly
assigned from a power law distribution scaling as $v^{-3.8}$).  For each
lump, we estimate its total flux as in the previous cases, integrating
Eq.~\ref{eq:rflux} with a density profile scaled according to 
Eq.~\ref{eq:consrule} for the concentration.

In the absence of further constraints on the possible value for ${\rm
R_{min}}$, we use the same criteria as before and choose it to be a fixed
fraction of the virial radius, ${\rm R_{min}} \approx 10^{-12}r_{200}$. 
This corresponds to a mean density of $\approx 10^{22}M_\odot$ kpc$^{-3}$  
for the galactic halo. The results are not too sensitive to the value of the
minimum integration radius as is apparent from inspection of Fig.~\ref{f:007}. The
total flux is then averaged over the entire sky and we repeat this
process in order to estimate the variance. The cumulative
distribution of flux above a given substructure peak circular velocity 
($\Sigma\Phi_{TOT}$) is plotted for ten of these random halo realisations in
Fig.~\ref{f:008}.

It is evident from this plot that the effects of including the entire mass
spectrum of substructure is quite dramatic and boosts the expected flux from
the smooth halo by several orders of magnitude. However, most of the flux
arises from the subhalos with circular velocities larger than 10 km s$^{-1}$.
Extrapolating to very small halos would not change the total flux by a large
factor.

To quantify the brightening of the background due to substructure, we have 
calculate the average flux due to all clumps with $v_{peak}>1$ km s$^{-1}$
within a spherical halo. 
The point plotted in Fig.~\ref{f:007} represents this contribution to the 
flux, where the error-bar is the 1$\sigma$ variation among the different 
Monte-Carlo models. From this plot we see that the flux due to substructure
is over two orders of magnitude brighter than 
the smooth background from a spherical halo. We note that one 
needs to observe a fairly large fraction of the sky ($>100$ square degrees) to 
ensure a significant number of clumps lie in the field of view.
(Also note that the variance at high peak circular velocities is due to the 
proximity of the largest few dark matter substructures, however, the mean 
total flux converges to similar values for each Monte-Carlo model.)

\subsection{The flux due to substructure in prolate and oblate halos}

Not only is the mean flux at a given position on the sky dominated by
substructure halos, the spatial distribution of flux across the sky
will be determined by the substructure. The convergence study by Ghigna 
{\it et al.} (Ref.~\cite{ghigna2000}) shows that substructure halos trace the 
global mass distribution of the halo. Therefore, we can use the N-body 
simulations to generate Monte-Carlo distributions of substructure halos and 
construct all-sky maps of the expected gamma-ray flux. We take a random 
particle from the simulation and assign a circular velocity from a 
distribution $dn(v_c)/dv_c \propto v_c^{-3.8}$. For each sub-halo we 
calculate its total annihilation flux and then repeat the process until we 
have 500,000 halos above a circular velocity of 1${\rm km s}^{-1}$.

Figure~\ref{f:009}(a) and Fig.~\ref{f:009}(b) show the resulting 
sky distribution of flux from sub-halos binned in one degree bins 
where the observer has been placed in the short and long axis of the 
simulation respectively. Large substructure halos, such as the Magellanic 
Clouds in our own halo, will contain its own gravitationally bound sub-halos 
which leads to clustering of gamma-ray emission in the all sky maps.

Future observations may only be able to make deep strip maps therefore in
Fig.~\ref{f:010} we have binned the flux along lines of constant galactic $l$
and $b$ with the observer placed in the short and long axis of the global
density distribution.  From these plots we can see that the emission 
from substructure peaks at the galactic center, as one would expect, this 
effect is not that different for spherical halos than for prolate or oblate 
halos. 

Since the substructure traces the global mass distribution, a prolate
halo would also have a prolate distribution of satellites. Therefore
we can study the variation of flux within smooth prolate or oblate
halos to examine how the background flux from substructure can be used
to quantify the halo shape. We calculate the observed flux as a
function $l$ and $b$ for spherical, prolate and oblate flattened 2:1
and 3:1 geometries.  In each case, the density profile is taken from
Eq.~\ref{eq:0} and again the observer is placed in either the short
(Figure~\ref{f:011b}(a)) or long (Figure~\ref{f:011b}(b)) axis of
symmetry.

These plots show how the distribution of flux on the sky can vary
significantly depending on the shape of the density distribution and 
on where the observer is situated within the halo.  

\subsection{The distribution of point sources}

Individual substructures may be observed and quantified if the resolution of
the telescope is sufficient. However, all of the past and present observations
would only detect substructure as unresolved point sources.  The distribution
of their fluxes (and spatial distribution on the sky) may be used to rule out
alternative origins, such as extra-galactic sources. In 
Fig.~\ref{f:012b} we plot the cumulative distribution of point sources above 
a given flux within one degree square bins.  The two curves consider 
substructure with peak circular velocities larger than 10 km s$^{-1}$ and 1 
km s$^{-1}$. The number density of the brightest sources in the sky scales as
$N\propto F^{-0.7}$.

Higher resolution simulations are vital to quantify how much substructure
survives within the galactic halo, how it is spatially distributed and to
quantify the internal structure of surviving substructure. However,
Fig.~\ref{f:012b} gives an idea of what to expect if an all sky survey is
carried out that is capable of detecting the brightest substructure halos.

\section{Conclusions}
\label{s:concs}

Numerical simulations that follow the growth of structure within a
universe dominated by neutralinos (cold dark matter) have achieved a
resolution that allows their global structure and internal structure
to be quantified.  The density profiles, shapes of dark matter halos,
abundance and properties of dark matter substructure, all play an
important role in determining the absolute surface brightness of
observable products from dark matter annihilation.

We have used the results from the highest resolution simulations ever
performed of CDM halos to examine the expected all-sky distribution of
gamma-rays from neutralino annihilation. Substructure can boost the
expected flux significantly over that originating from a smooth dark
matter halo.  Thus, gamma-ray observations, such as EGRET data, may
already have the potential of constraining a large part of the
parameter range of the neutralino cross-sections.  The distinguishing
shapes of CDM halos and the unique spatial and flux distribution of
point sources from substructure within the Galactic halo should allow
a unique identification of observational data with dark matter.

\section*{Acknowledgments}

The authors would like to thank Prof. Arnold Wolfendale for numerous
discussions and suggestions that have improved the quality of this
work.  Carlos Calc\'{a}neo--Rold\'{a}n continues his research thanks
to the generous support from the People of M\'{e}xico, through a grant
by CONACyT.  Ben Moore is a Royal Society research
fellow. Computations were carried out as part of the Virgo consortium.

\begin{figure}
\caption{ (a) The circular velocity curves $V_c(r) = \sqrt{GM(r)/r}$, and (b)
density profiles are plotted as a function of the radius for each of
the halo models considered in the text.}
\label{f:001} 
\end{figure}

\begin{figure}
\caption{The gamma ray flux from neutralino annihilation, $\phi(\psi)$, 
plotted as a function of the angular distance from the galactic center 
$\psi$. The curves show the results using the three different density 
profiles plotted in Fig.~\ref{f:001} The flux at a given position is averaged
over $4\pi$ steradians.}
\label{f:002} 
\end{figure}

\begin{figure}
\caption{All-sky maps of the gamma ray background constructed using a single 
high-resolution N-body simulation of a cold dark matter halo. The observer 
has been placed in the short (a) and long (b) axis of the simulated halo.}
\label{f:003} 
\end{figure}

\begin{figure}
\caption{The left panel shows a unit oblate ellipsoid and the right hand 
panel shows a unit prolate ellipsoid. The axial ratios for both are 2:1.}
\label{f:004} 
\end{figure}

\begin{figure}
\caption{The gamma ray flux, $\phi$, plotted as a function of angle $\psi$, 
for smooth halos of the same total mass using the density profile given in 
Eq.~\ref{eq:0} for spherical, oblate and prolate halo geometries. The points 
are values of the flux measured directly from the N-body halo illustrated in 
Fig.~\ref{f:003}.}
\label{f:005} 
\end{figure}

\begin{figure}
\caption{A sketch showing the geometry of an observer in the galaxy 
viewing substructure in the galactic halo.}
\label{f:006} 
\end{figure}

\begin{figure}
\caption{The gamma ray flux, $\Phi_{AV}$, plotted as a function of minimum 
integration radius R$_{{\rm min}}$ for halo substructure of different circular
velocities and distances as detailed in the text. The shaded region 
shows the range of background values at the Galactic anti-center that can be 
expected depending on the halo shape. The point is the average flux due to 
all clumps with $v_{peak}>1$ km s$^{-1}$. Note that the size of the error bar on
this point depends on the area of the sky surveyed.
} 
\label{f:007}
\end{figure}

\begin{figure}
\caption{The cumulative gamma-ray flux from halo substructures, 
$\Sigma \phi_{TOT}(v>v_{peak})$, above a given substructure circular velocity
$v_{peak}$. The ten different curves correspond to different 
Monte-Carlo realizations of a Galactic halo of substructure halos.
The flux is averaged over $4\pi$ steradian and can be compared with
the flux from the smooth halo from Fig.~\ref{f:002} and Fig.~\ref{f:005}.}
\label{f:008}
\end{figure}

\begin{figure}
\caption{All-sky map of the gamma ray background that arises solely from
dark matter substructures. The positions and circular velocities of sub-halos
above a circular velocity of 1${\rm km s}^{-1}$ are drawn from the N-body 
simulations but the flux from each halo is calculated analytically. 
The observer is located on the short (a) and long (b) axis of symmetry.
The grey scale corresponds to the log of the flux of annihilation products.}
\label{f:009} 
\end{figure}

\begin{figure}
\caption{The average gamma-ray flux per square degree from dark matter 
substructure as measured within the simulated CDM halo
along a great circle of constant galactic latitude 
(a) and longitude (b). The average has been taken over a strip of width $44$ 
degrees, the left hand plot represents the view along the 
short axis while the right hand side is the view along the long axis.}
\label{f:010} 
\end{figure}

\begin{figure}
\caption{The effect of halo shape on the gamma-ray flux. Halo density profiles are 
drawn from spherical, oblate or prolate distributions with the indicated 
axis ratios. The observer is placed in the short axis (a) whilst the in 
(b) the observer is in the long axis.}
\label{f:011b} 
\end{figure}

\begin{figure}
\caption{The cumulative number of gamma-ray sources above a 
given flux within a window $\Delta\Omega = 1^{\circ} \times 1^{\circ}$. 
The two curves are for substructure halos with circular velocities 
larger than $10 {\rm km s^{-1}}$ (dashed line) and $1 {\rm km s^{-1}}$ 
(solid line). }
\label{f:012b} 
\end{figure}

\clearpage

\begin{figure}
\centering
\epsfxsize=\hsize\epsffile{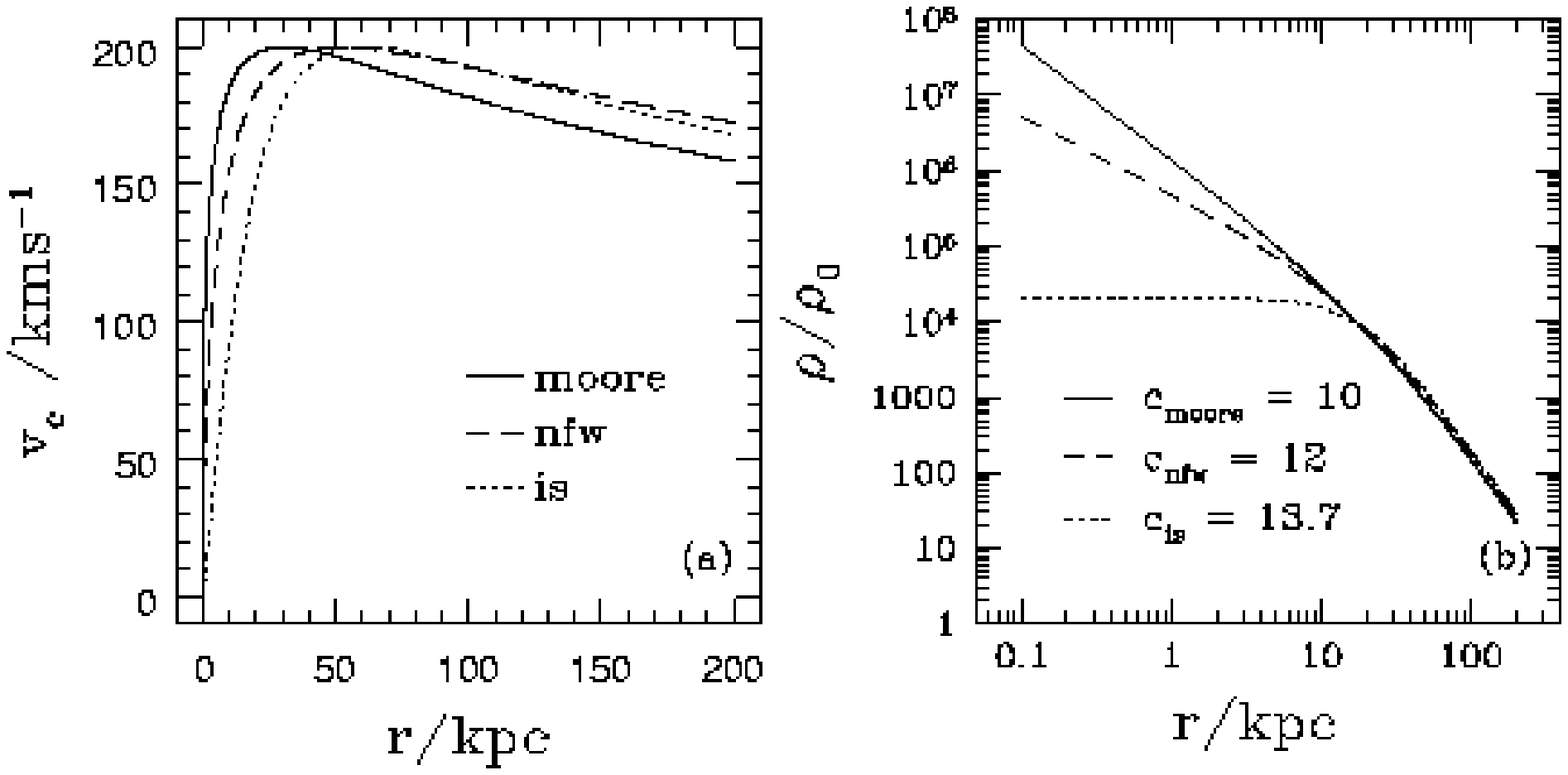}
\end{figure}

~\\

~\\

Figure ~\ref{f:001}

\begin{figure}
\centering
\epsfxsize=\hsize\epsffile{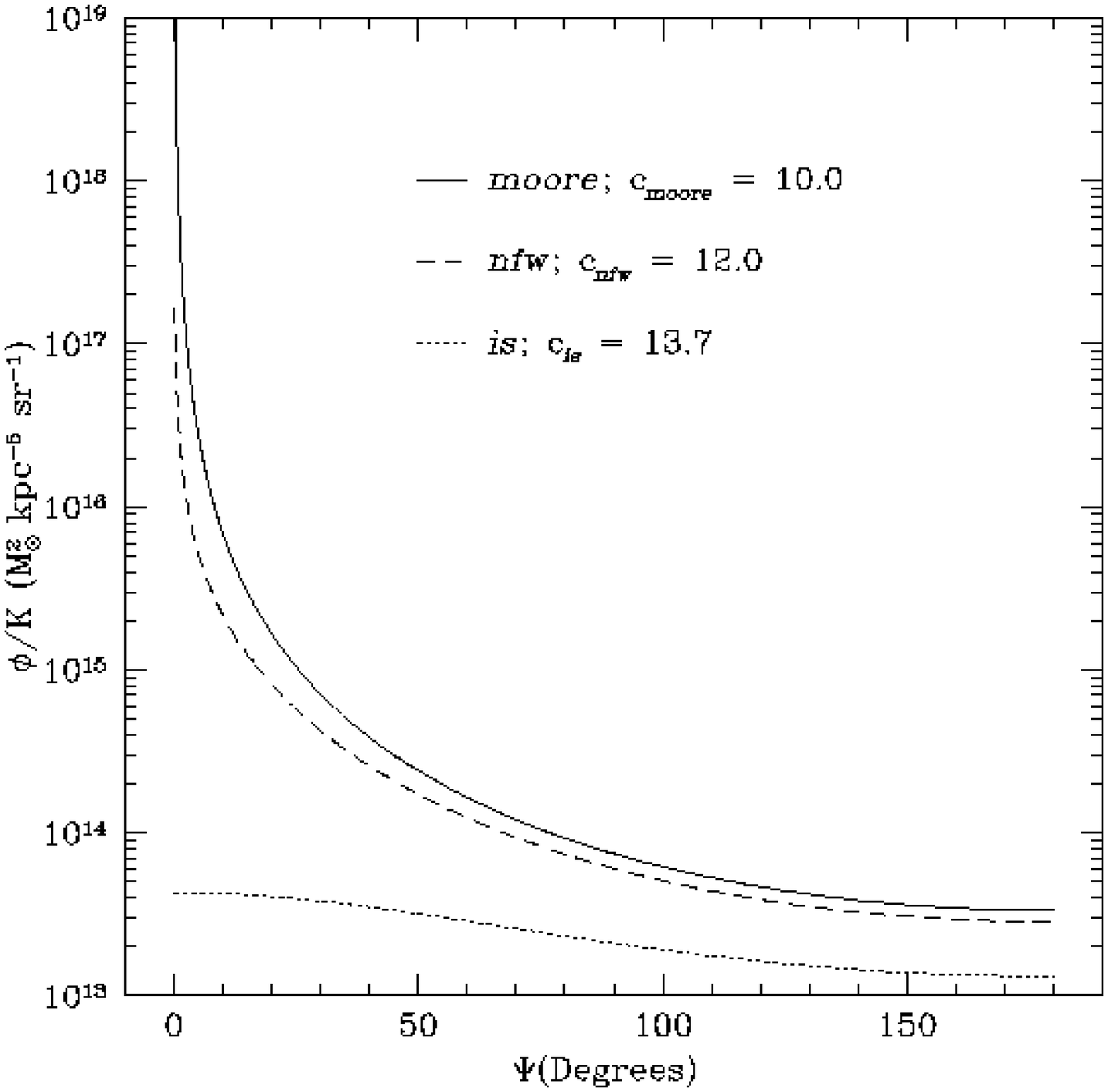}
\end{figure}

~\\

~\\

Figure ~\ref{f:002}

\begin{figure}
\centering
\epsfxsize=\hsize\epsffile{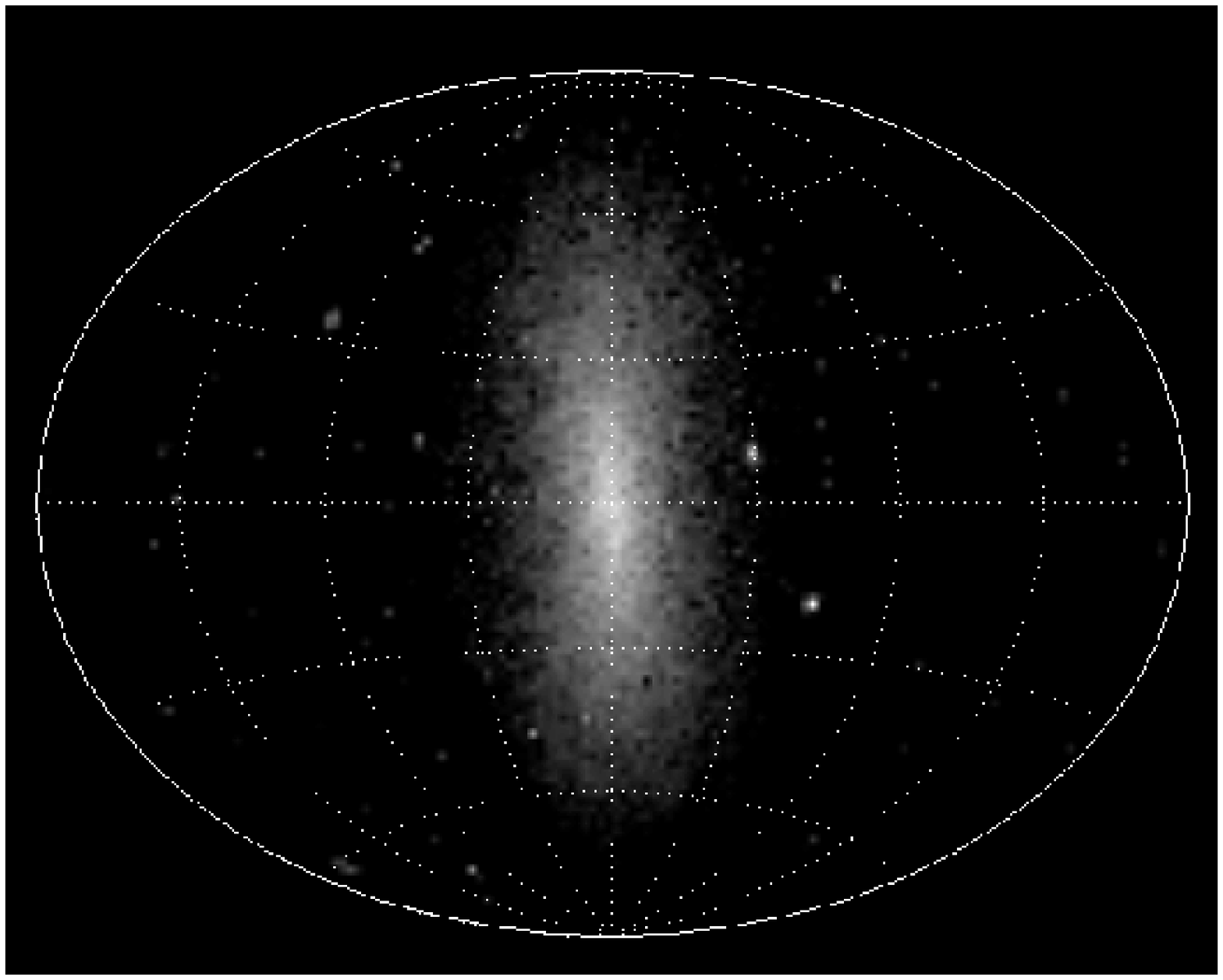}
\end{figure}

~\\

~\\

Figure ~\ref{f:003}(a)

\begin{figure}
\centering
\epsfxsize=\hsize\epsffile{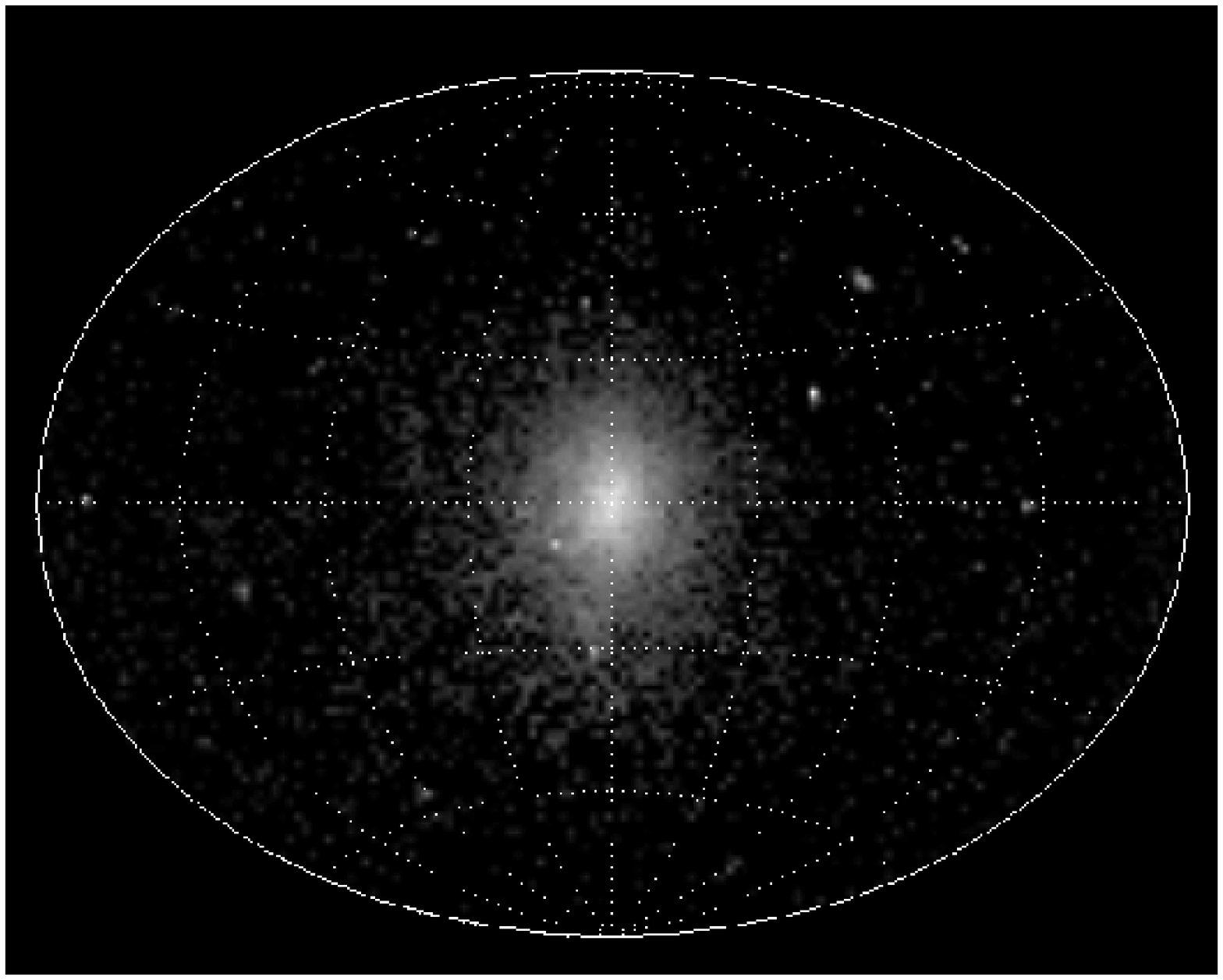}
\end{figure}

~\\

~\\

Figure ~\ref{f:003}(b) 

\clearpage
\begin{figure}
\centering
\epsfxsize=\hsize\epsffile{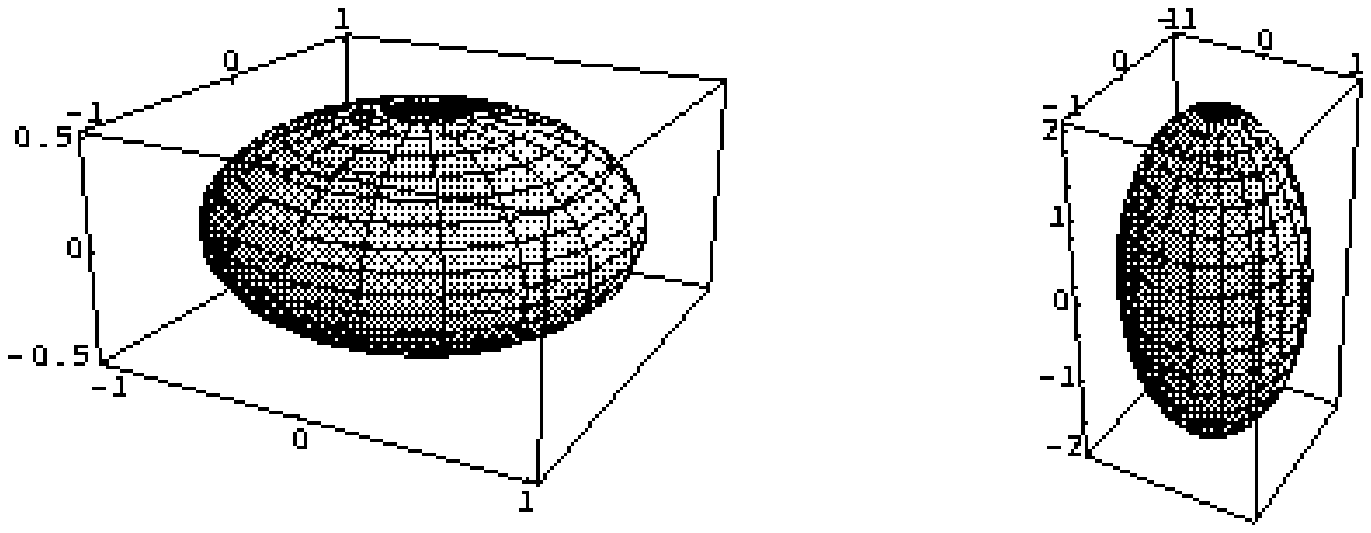}
\end{figure}

~\\

~\\

Figure ~\ref{f:004}

\begin{figure}
\centering
\epsfxsize=\hsize\epsffile{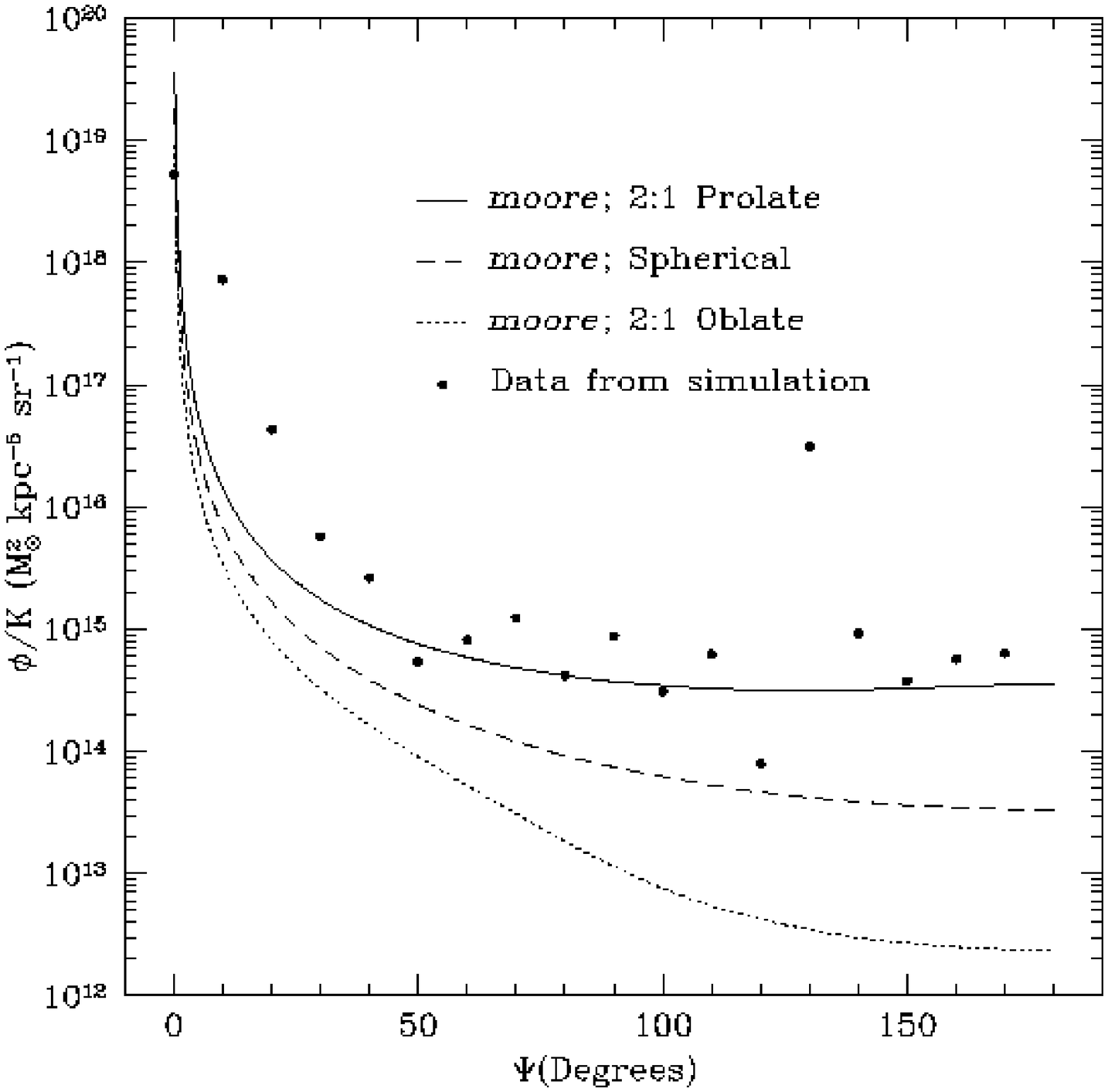}
\end{figure}

~\\

~\\

Figure ~\ref{f:005}

\begin{figure}
\centering
\epsfxsize=\hsize\epsffile{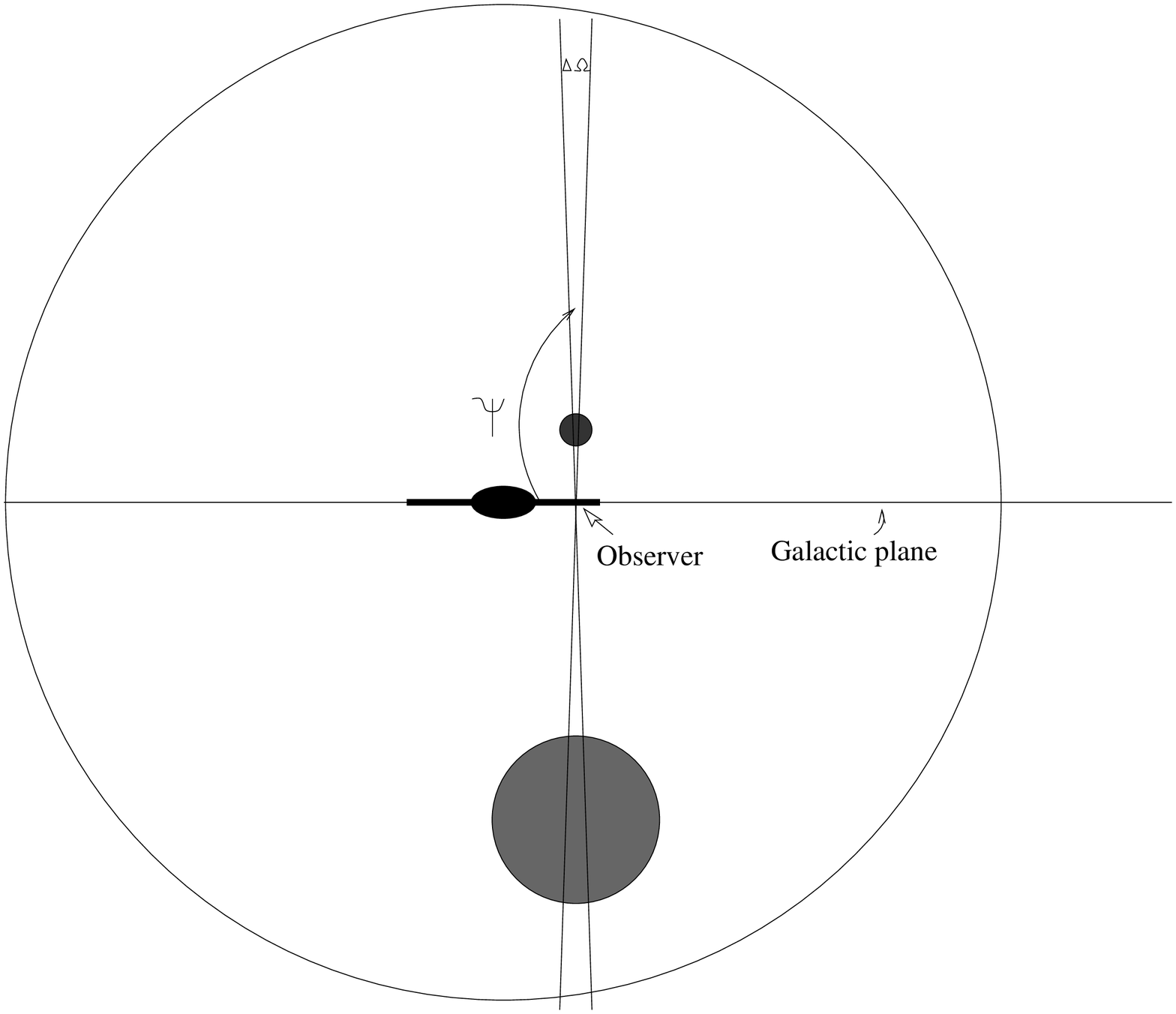}
\end{figure}

~\\

~\\

Figure ~\ref{f:006} 

\begin{figure}
\centering
\epsfxsize=\hsize\epsffile{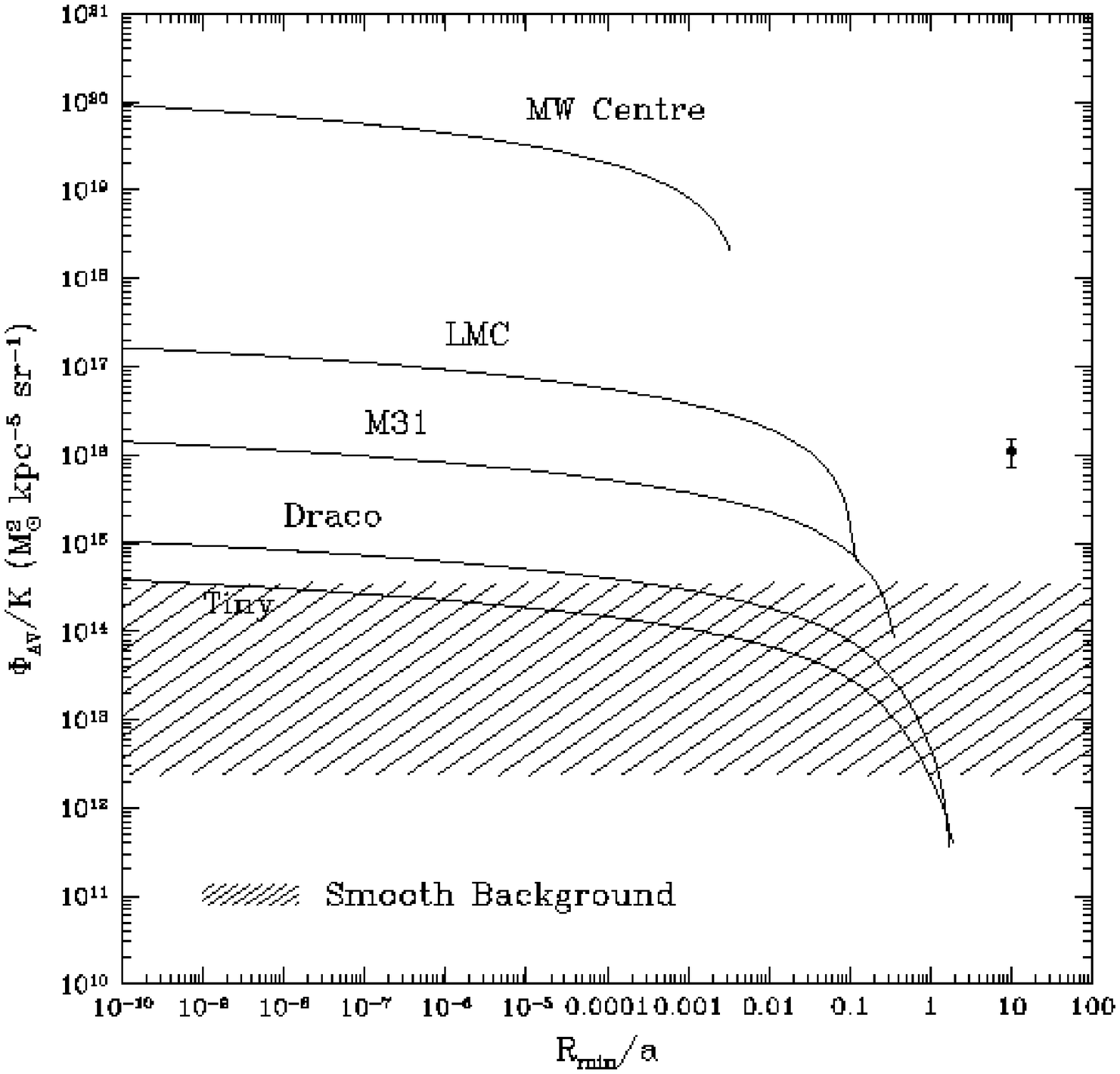}
\end{figure}

~\\

~\\

Figure ~\ref{f:007}

\begin{figure}
\centering
\epsfxsize=\hsize\epsffile{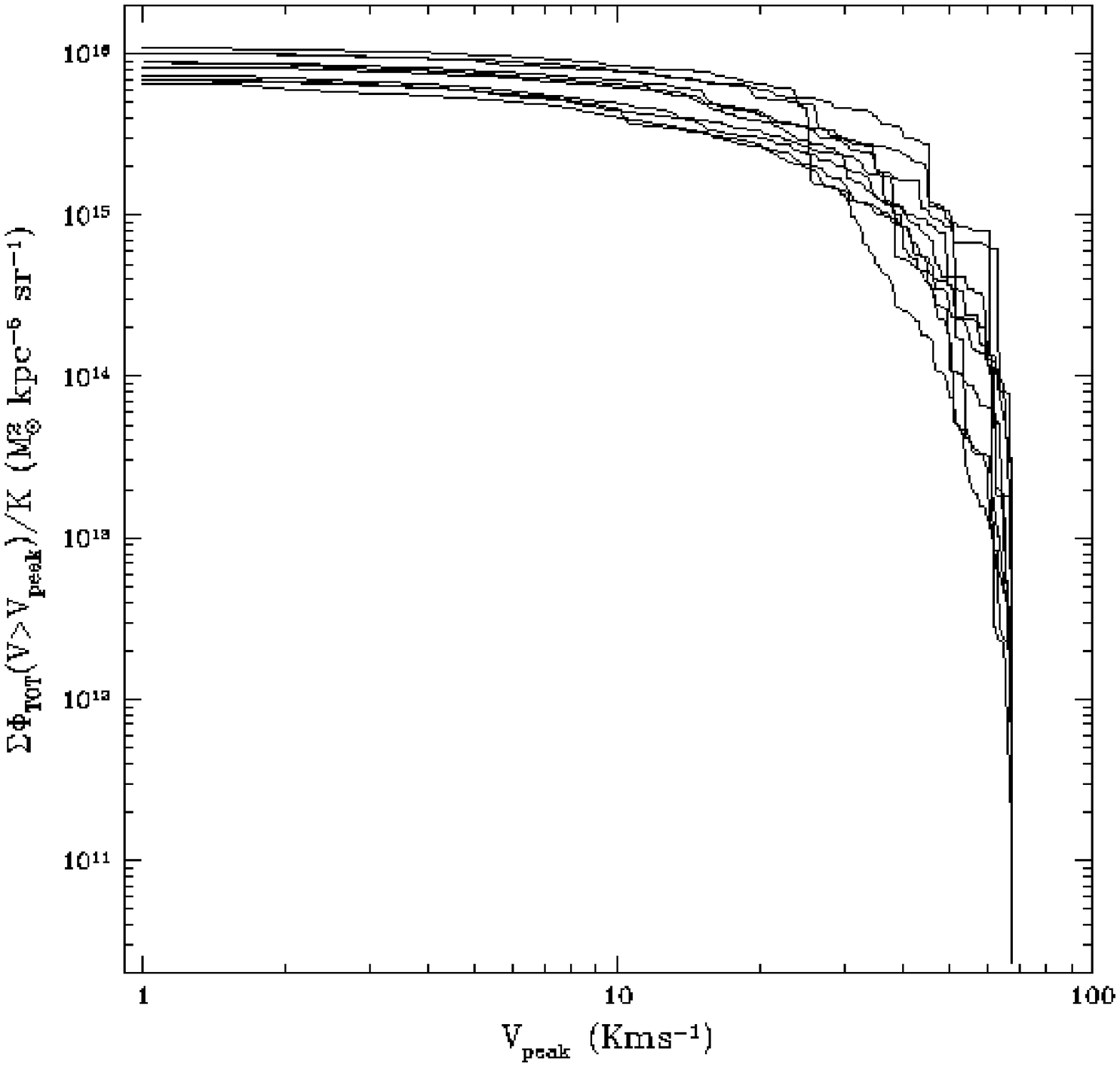}
\end{figure}

~\\

~\\

Figure ~\ref{f:008} 

\begin{figure}
\centering
\epsfxsize=\hsize\epsffile{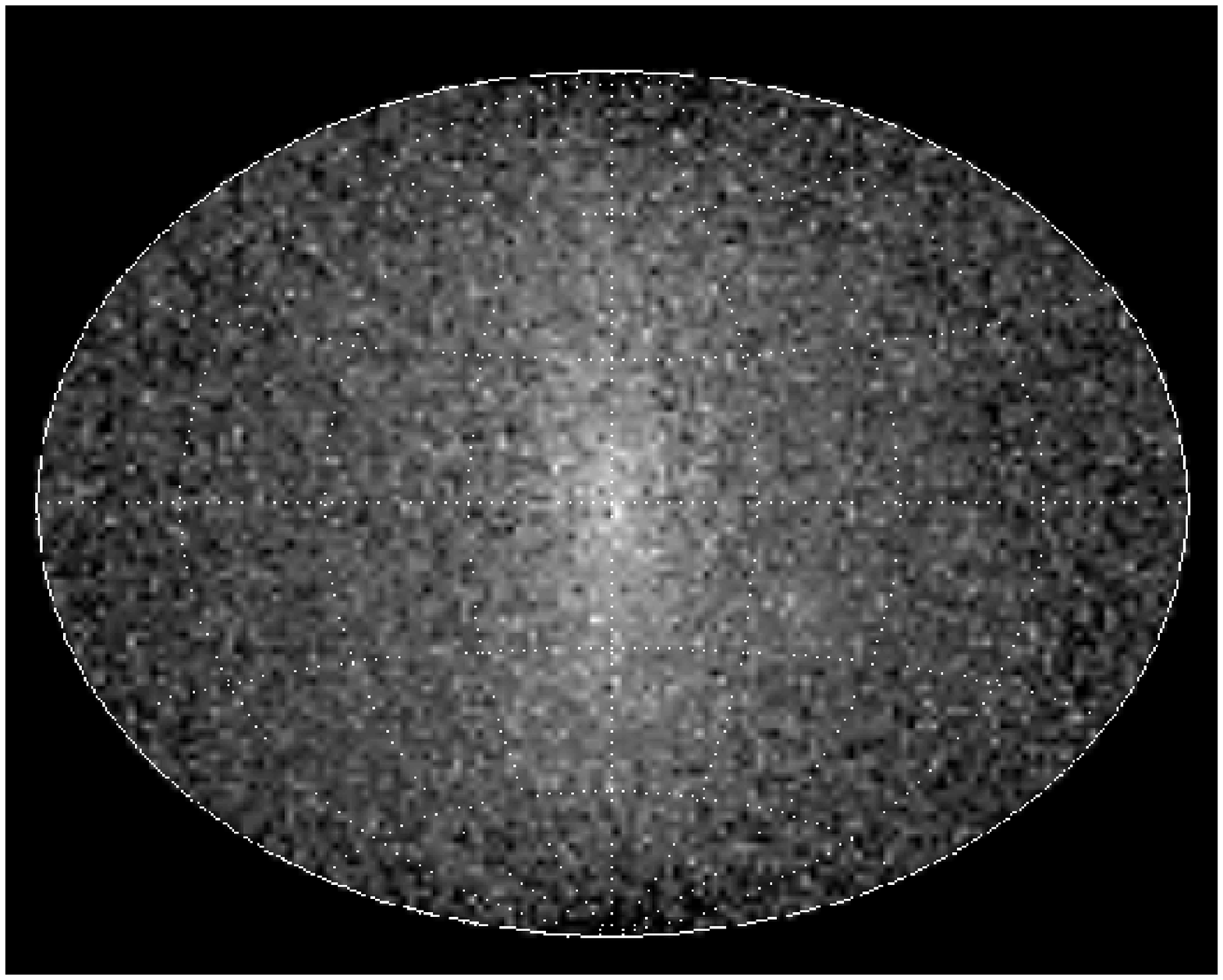}
\end{figure}

~\\

~\\

Figure ~\ref{f:009}(a)

\begin{figure}
\centering
\epsfxsize=\hsize\epsffile{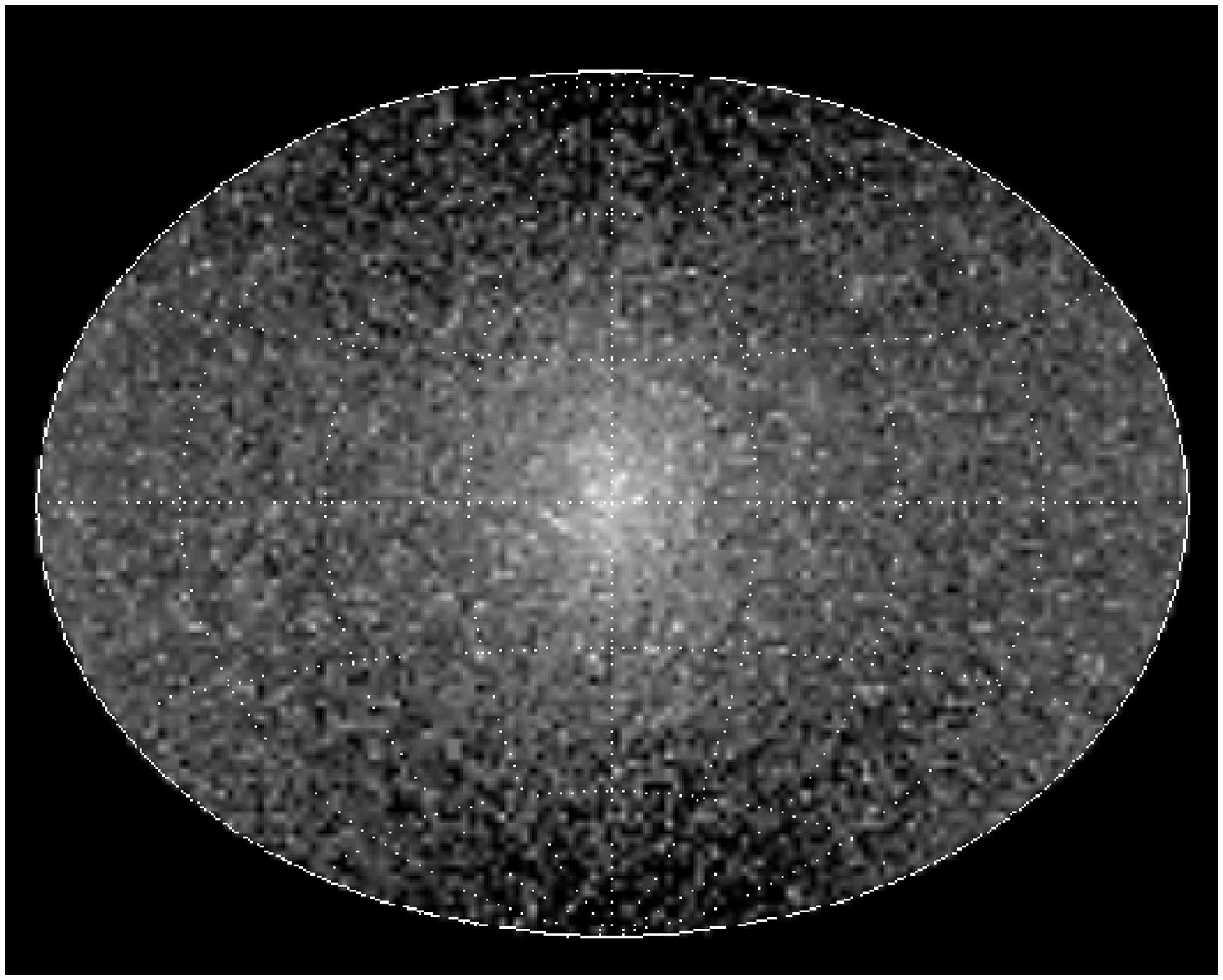}
\end{figure}

~\\

~\\

Figure ~\ref{f:009}(b) 

\begin{figure}
\centering
\epsfxsize=\hsize\epsffile{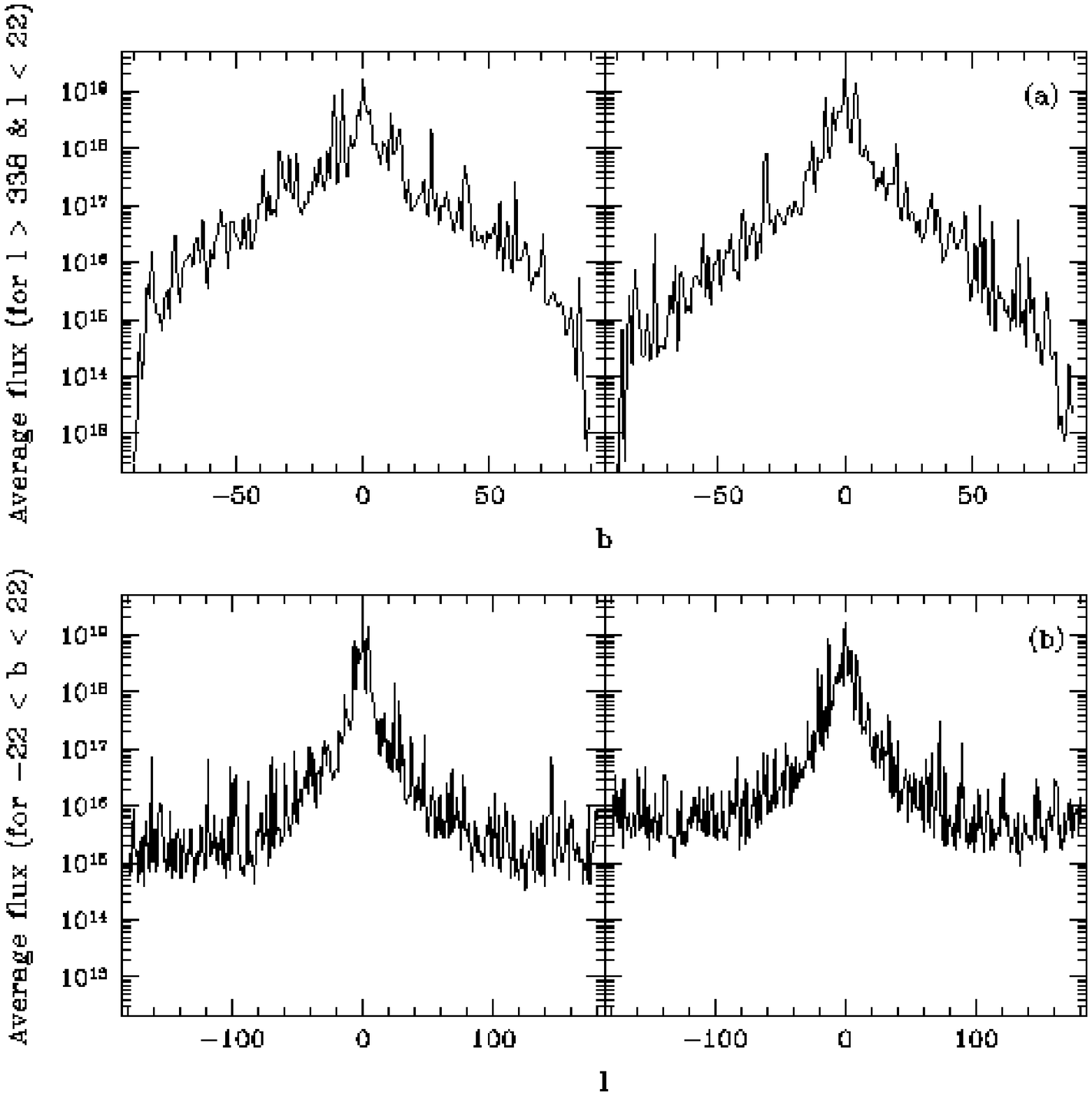}
\end{figure}

~\\

~\\

Figure ~\ref{f:010} 

\begin{figure}
\centering
\epsfxsize=\hsize\epsffile{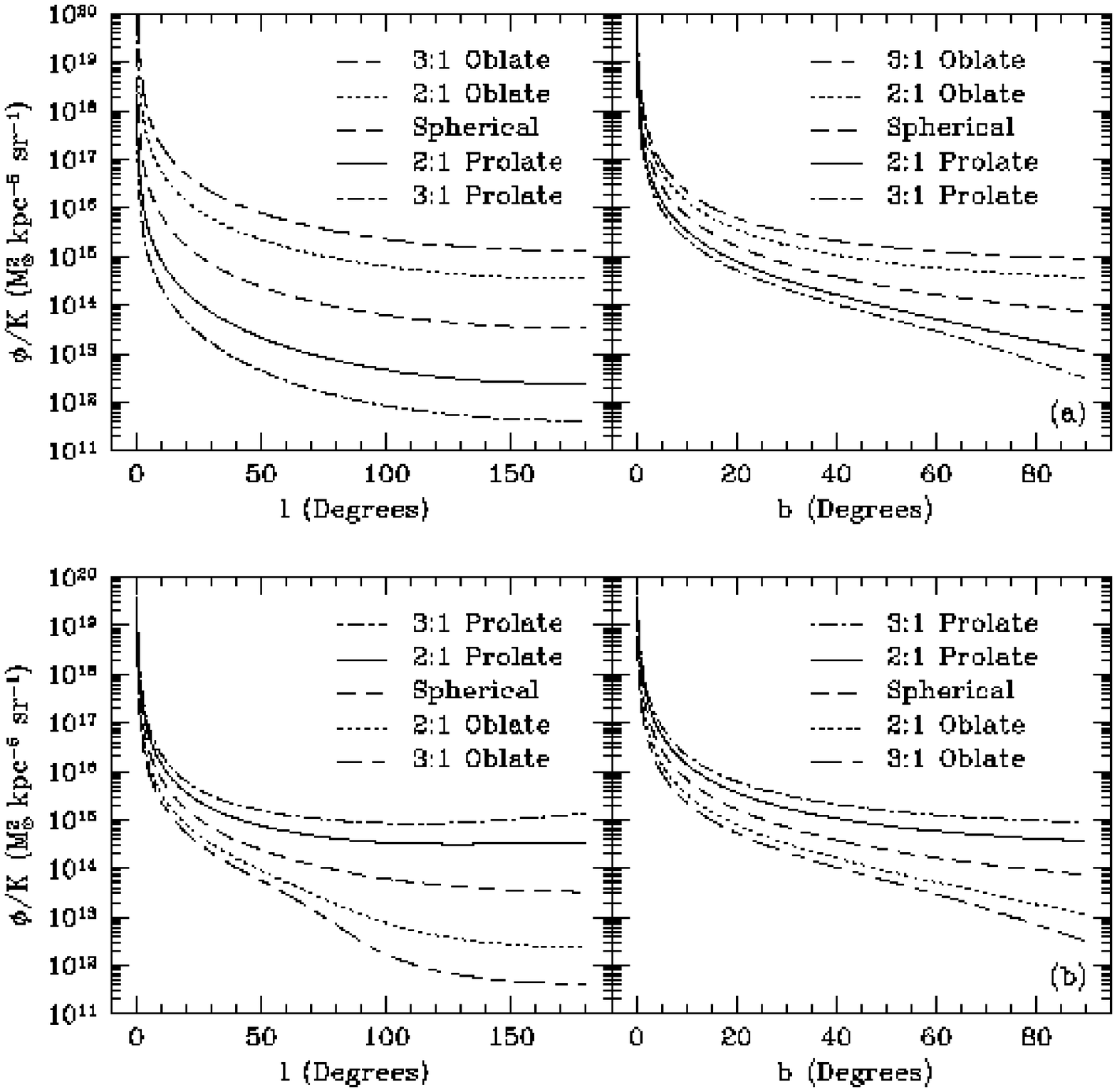}
\end{figure}

~\\

~\\

Figure ~\ref{f:011b} 

\begin{figure}
\centering
\epsfxsize=\hsize\epsffile{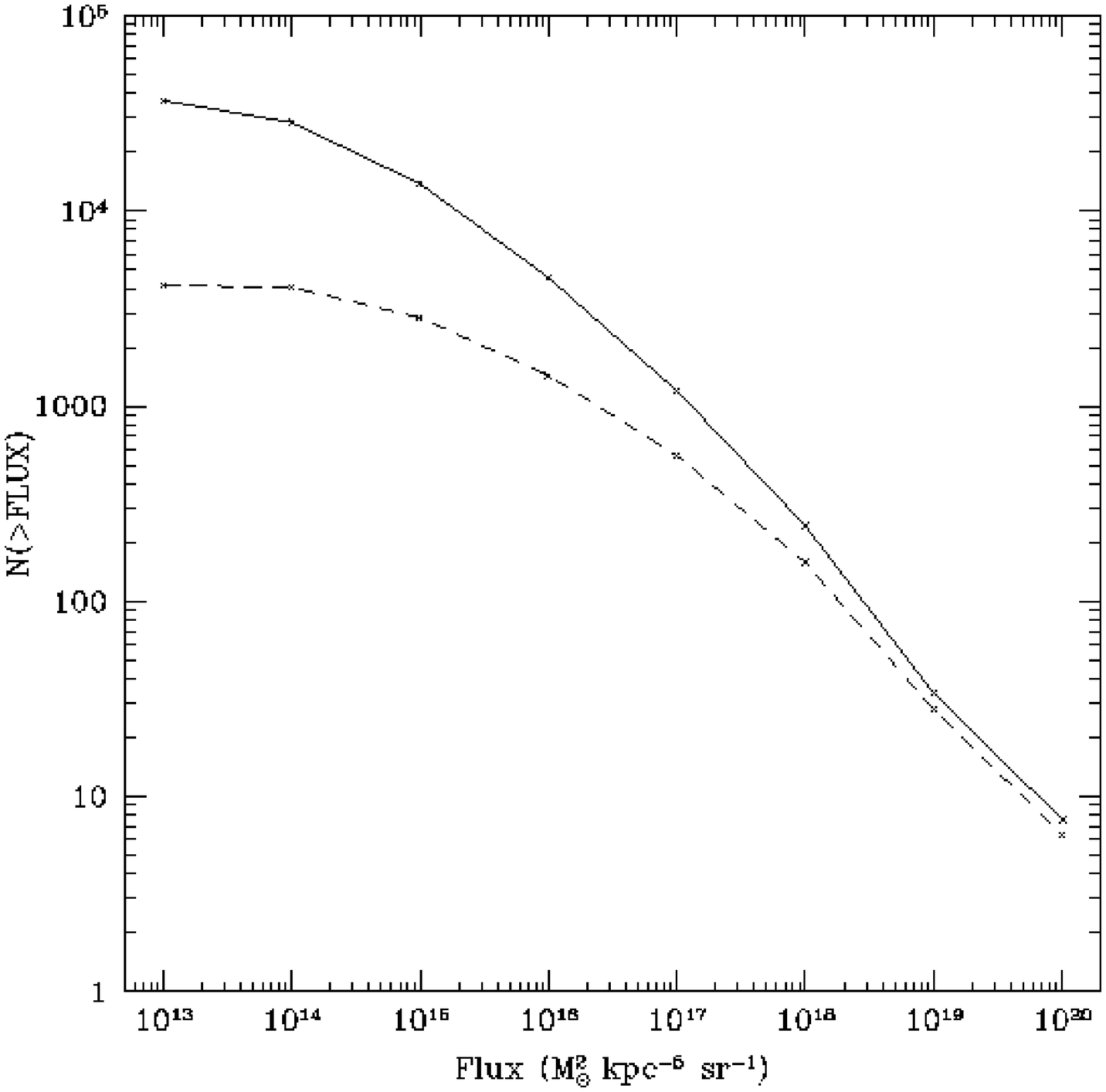}
\end{figure}

~\\

~\\

Figure ~\ref{f:012b}

\end{document}